\documentclass{aa}
\usepackage{graphicx}
\usepackage{natbib}
\usepackage{txfonts}
\usepackage{subfigure,xtab}
\usepackage{lscape}

\bibpunct{(}{)}{;}{a}{,}{,}

\newcommand{\nodata}{---}

\usepackage{fancybox}
\usepackage{booktabs}

\begin{document}


\title{Follow-up of 27 radio-quiet gamma-ray pulsars 
at 110--190 MHz using the international LOFAR station FR606}

\author{J.--M. Grie{\ss}meier
          \inst{1,2}
          \and
          D. A. Smith
          \inst{3,4}
          \and
          G. Theureau
          \inst{1,2,5}
          \and
          T. J. Johnson
          \inst{6}
          \and
          M. Kerr
          \inst{7}
          \and
          L. Bondonneau
          \inst{1,8}
          \and
          I. Cognard
          \inst{1,2}
          \and
          M. Serylak
                \inst{9,10}
 }

\institute{
        LPC2E - Universit\'{e} d'Orl\'{e}ans / CNRS, France. \email{jean-mathias.griessmeier@cnrs-orleans.fr}
        \and
Station de Radioastronomie de Nan\c{c}ay, Observatoire de Paris - CNRS/INSU, USR 
  704 - Univ. Orl\'{e}ans, OSUC, route de Souesmes, 18330 Nan\c{c}ay, France
        \and
        Centre d’\'{E}tudes Nucl\'{e}aires de Bordeaux Gradignan, IN2P3/CNRS, Universit\'{e} Bordeaux, 33175 Gradignan, France
        \and
        Laboratoire d'Astrophysique de Bordeaux, Universit\'e Bordeaux, B18N, all\'ee Geoffroy Saint-Hilaire, 33615 Pessac, France
        \and
        Laboratoire Univers et Th\'eories LUTh, Observatoire de Paris, CNRS/INSU, Universit\'e Paris
Diderot, 5 place Jules Janssen, 92190 Meudon, France
        \and
        College of Science, George Mason University, Fairfax, VA 22030, resident at Naval Research Laboratory, Washington, DC 20375, USA
        \and
        Space Science Division, Naval Research Laboratory, Washington, DC 20375-5352, USA
        \and
        LESIA, Observatoire de Paris, CNRS, PSL, SU/UP/UO, 92195 Meudon, France
        \and
        South African Radio Astronomy Observatory, 2 Fir Street, Black River Park, Observatory 7925, South Africa
        \and
        Department of Physics and Astronomy, University of the Western Cape, Bellville, Cape Town 7535, South Africa
}

\date{Version of \today}

\abstract{The \textit{Fermi} Large Area Telescope has detected over 260 gamma-ray pulsars. 
About one quarter of these are labeled as radio-quiet,
that is~they either have radio flux densities $<30\,\mu$Jy at 1400 MHz, 
or they are not detected at all in the radio domain. 
In the population of nonrecycled gamma-ray pulsars, the fraction of radio-quiet pulsars is higher, about one half.
}
{
Most radio observations of gamma-ray pulsars have been performed at frequencies between 300 MHz and 2 GHz. 
However, pulsar radio fluxes increase rapidly with decreasing frequency, and their radio beams often broaden at low frequencies.
As a consequence, some of these pulsars might be detectable at low radio 
frequencies even when no radio flux is detected above 300 MHz.
Our aim is to test this hypothesis with low-frequency radio observations.
}
{We have observed 27 \textit{Fermi}-discovered gamma-ray pulsars with the international LOw
Frequency ARray (LOFAR) station FR606 in single-station mode. 
We used the LOFAR high band antenna (HBA) band (110--190 MHz), with an average observing time of 13\,h per target.
Part of the data had to be discarded due to radio frequency interference. 
On average, we kept 9\,h of observation per target after the removal of affected datasets, 
resulting in a sensitivity for pulse-averaged flux on the order of 1--10 mJy.
}
{
We do not detect radio pulsations from any of the 27 sources, 
and we establish stringent upper limits on their low-frequency radio fluxes. 
These nondetections are compatible with the upper limits derived from 
radio observations at other frequencies.
We also determine the pulsars' geometry from the gamma-ray profiles
to see for which pulsars the low-frequency radio beam is expected to cross Earth.
}
{
This set of observations provides the most constraining upper limits on 
the flux density at 150 MHz for 27 radio-quiet gamma-ray pulsars. 
In spite of the beam-widening 
expected at low radio frequencies,
most of our nondetections can be explained by an unfavorable viewing geometry;
for the remaining observations,
especially those of pulsars detected at higher frequencies,
the nondetection is compatible with 
insufficient sensitivity.
}


\titlerunning{Follow-up of 27 radio-quiet gamma-ray pulsars with LOFAR FR606}
\authorrunning{J.--M. Grie{\ss}meier et al.}

\maketitle

\section{Introduction}

The Large Area Telescope (LAT) on the \textit{Fermi} satellite has 
strongly increased the number of known gamma-ray pulsars 
\citep[see e.g., the ``\textit{Fermi} 2nd Pulsar Catalog''][hereafter 2PC]{Abdo13}. 
A major update, ``3PC,'' which is in preparation, will characterize at least 260 gamma-ray pulsars\footnote{A list is maintained at https://confluence.slac.stanford.edu/display/GLAMCOG/Public+List+of+LAT-Detected+Gamma-Ray+Pulsars}. 
For the \textit{Fermi} LAT pulsar detections, three different approaches have been used \citep{Pletsch12_755}: 
(1) For some pulsars known from radio observations, gamma-ray pulsations have been detected using radio ephemerides, 
see for example \citet{Smith19}. 
(2) Radio searches of unassociated gamma-ray sources have revealed new radio pulsars, mostly millisecond pulsars (MSPs). 
A radio ephemeris could be then derived, allowing for the detection of gamma-ray pulsations.
(3) Data from other unassociated gamma-ray sources have been searched 
for a large number of trial parameters (e.g., period, period derivative, position, among others),  
allowing for the detection of new ``blind search'' gamma-ray pulsars. 
In a half-dozen cases,
radio follow-up and folding at the gamma-ray derived period identified these sources 
as radio pulsars \citep{Clark18}.
Some of these are very faint: In 2PC, a pulsar is designated as radio-quiet 
(RQ)
if its flux density at 1400 MHz, $S_{1400}$, is $<30\,\mu$Jy,
for which 
J1907+0602 with $S_{1400} = 3.4\,\mu$Jy is an example \citep{Abdo10_711}.
In the majority of cases, however, no radio counterpart has been detected.  
Approximately half of the 142 nonrecycled gamma-ray pulsars are radio-loud
($S_{1400} \ge30\,\mu$Jy), the rest are radio-quiet \citep[][]{Wu18}. 

If detected, radio emission from a gamma-ray pulsar provides extra information, such as
the pulsar's dispersion measure (DM). Combined 
with a geometrical model of the Galaxy 
\citep[e.g.,][]{Cordes02,Yao17}, this provides an approximate source distance.
This 
has motivated a number of radio follow-up studies for known gamma-ray 
pulsars\footnote{For the studies mentioned in the following, 
measured flux density values or upper limits thereof are given in 
Appendix \ref{sec-appendix} for the pulsars that overlap with our selection.}.

Most of these 
follow-up studies 
have been performed between 820 and 1500 MHz. 
Studies at lower frequencies are interesting for two reasons.
First,
the spectra of most pulsars are usually well described by 
a power law, 
but different pulsars have different spectral indices 
\citep{Sieber73,Maron00,Jankowski18}.
Some of the gamma-ray pulsars might have a steeper than average spectrum 
with a very low (possibly undetected) flux density at high frequencies, 
but a high (and potentially detectable) flux at lower frequencies. 
Second, the radio beams of many pulsars are wider at low frequencies
\citep[e.g.,][]{Sieber75}, 
such that a beam that narrowly misses Earth at higher frequencies
could still intersect Earth at lower frequencies.
Such beam widening is expected, for example, from the radius-to-frequency mapping model 
\citep[i.e., higher frequency emission originating closer to the neutron star than lower frequency emission,][]{Cordes78}.

With beam widening in mind, \citet{Maan14} performed 
follow-up observations 
of \textit{Fermi} LAT detected pulsars
at 34 MHz using the Gauribidanur telescope.
However, many pulsars show a spectral break or turnover between 50 and 100 MHz
\citep[e.g.,][]{Sieber73,Kuzmin78,Izvekova81},
so that observations at 34 MHz may miss pulsars which might be detectable at frequencies above this spectral turnover.
In addition, scatter broadening may lead to a pulse width exceeding the pulse period, making
detection difficult. Scatter broadening is usually assumed to scale as $\propto f^{-4}$, where $f$ is the observing frequency, so that low-frequency observations are particularly strongly impacted.
Indeed, \citet{Maan14} did not detect any periodic signal above 
their detection threshold ($8\,\sigma$)
at 34 MHz, but they were able to determine flux density upper 
limits.

Flux density upper limits at 150 MHz were derived for all observable 
radio-quiet gamma-ray pulsars using the GMRT all-sky survey TGSS 
\citep{Frail16}.\ However,
the integration time per pointing was only 15 minutes, 
leading to relatively weak constraints on the 
flux density.

For these reasons, 
we systematically followed up on all 27 northern sky 
nonrecycled radio-quiet gamma-ray pulsars with targeted observations 
and searched for radio pulsations at low frequencies.
For this, we used the international LOw
Frequency ARray (LOFAR) station FR606 in Nan\c{c}ay in stand-alone mode, observing in the 
high band antenna (HBA) frequency range (110--190 MHz). 

This paper is organized as follows:
Section \ref{sec-motivation} discusses the interest of low-frequency observations
(higher flux at low frequencies in Subsection \ref{sec-spectral-index} 
and beam widening at low frequencies in Subsection \ref{sec-beam-widening}).
Section \ref{sec-methods} describes the methods used:
We describe how the targets were selected (Subsection \ref{sec-target-selection}),
how the observations were performed (Subsection \ref{sec-observations}),
how the data were processed (Subsection \ref{sec-data-analysis}),
and how flux density upper limits were calculated (Subsection \ref{sec-flux-limit}).
In Section \ref{sec-results}, we present the results of our observations.
In Section \ref{sec-results-beam-width}, we discuss 
the pulsars' geometry:
We first discuss the consequences of beam widening for a uniform distribution
of pulsar geometries (Subsection \ref{sec-results-beam-width-motivation}),
then explain how we determined the pulsars' geometry from the observed gamma-ray profiles
(Subsection \ref{sec-results-beam-width-method}), and finally 
discuss the consequences of beam widening
for those specific geometries (Subsection \ref{sec-results-beam-width-discussion}).
Section \ref{sec-conclusions} closes with some concluding remarks.

\section{Radio beam basics} \label{sec-motivation}

\subsection{Radio spectra} \label{sec-spectral-index}

At frequencies $\gtrsim$200 MHz, the spectra of most pulsars are well described by 
a single spectral index \citep{Sieber73,Maron00,Jankowski18}.
However, different pulsars have different spectral indices, 
with a 
wide spread.
Throughout this work, we adopt an average spectral index of $-1.6$ 
with a standard deviation of 0.54 \citep{Jankowski18}. 
We assume that the spread of spectral indices is at 
least partially physical; for this reason, we use their standard deviation
rather than their standard error.
This range of values formally encompasses those given by other works, such as
\citet{Maron00} and \citet{Lorimer95}.
At frequencies below 200 MHz, the spectral index flattens
and most nonrecycled pulsars show a turnover between 50 and 100 MHz
\citep[e.g.,][]{Sieber73,Kuzmin78,Izvekova81,Bilous16,Jankowski18,Bilous20,Bondonneau20fr606}.
This is not accounted for in our calculations,
and the extrapolated flux density limits at frequencies $<200$ MHz 
are probably slightly overestimated.

With a spectral index of $-1.6 \pm 0.54$, we can expect flux densities 
to be higher at 150 MHz than at 1400 MHz by 1--2 orders of magnitude.
This gain is partially offset by the increased sky temperature 
against which the pulsar has to be detected; even so, for a comparable 
telescope gain, or effective area, we can expect to detect fainter pulsars.
This is particularly true for those that have a steeper than average spectral index.

No matter how steep the spectrum may be, a pulsar that is intrinsically faint or too 
distant remains
undetectable. 
A consensus about pulsar radio luminosity $L_r$ does not exist,
but it is frequently assumed 
that it depends on the rotation period ($P$) and period derivative ($\dot{P}$) as 
$L_r = L_0 P^a\dot P^b$, albeit with large uncertainties on the parameters $a$ and $b$ 
and with a large spread, that is to say $P^a\dot P^b$ and
$L_r$ are poorly correlated.
As an example, \citet{Johnston17} found $a,b$ such that $L_r = \dot E^{1/4}$ where
$\dot E$ is the spindown power.
Distance $d$ affects detectability in two ways: First, $S_{1400} \propto L_r/d^2$. Second, a large $d$
implies a large dispersion measure, especially at low Galactic latitudes.
For pulsars with periods $\lesssim 500$ ms and observations at
150 MHz, we are limited to DM $\lesssim 500$ pc cm$^{-3}$; for higher DM values, 
the expected scattering time exceeds the pulsar's rotation period (see Section \ref{sec-data-analysis}).
Thus some pulsars
are not seen even if the spectrum is steep and/or the 
beam broadens into the line of sight, as discussed below.

\subsection{Beam widening} \label{sec-beam-widening}

A simple radio beam model suffices for this work.
Following \citet{Cordes78} or, equivalently, the Pulsar Handbook \citep[][Section 3.4]{LorimerKramer05}, 
radio emission is centered around the neutron star magnetic dipole axis.
The dipole is inclined with an angle $\alpha$ from the rotation axis, and the line of sight from Earth makes an angle $\zeta$
with the rotation axis. High-energy electrons follow the magnetic field lines and radiate at radio frequencies that depend
on the field line curvature. Hence, radio emission is a cone-shaped beam of radius 
\begin{equation}
 \rho \approx \sqrt{{9\pi r_{em} \over 2cP}}  = \frac{3}{2} \sqrt {{r_{em} \over r_\text{LC} }}, 
 \label{rhoEq}
\end{equation}
where $r_{em}$ is the height of radio emission and 
$r_\text{LC} = {c P}/{(2\pi)}$ 
defines the ``light cylinder,'' 
the radius at which an object co-rotating with the neutron star with a spin period $P$ would reach the speed of light, $c$. 
Typical values of $r_{em}=300$\nolinebreak{} km and $P=0.1$ s give $ \rho \approx 20^\circ$. 
The radio beam sweeps Earth only if $(\alpha +  \rho) > \zeta > (\alpha - \rho)$, making the pulsar potentially radio-loud (RL).
Following \citet{Gil02}, the observed pulse has a width $W$ given by
\begin{equation}
        \sin^2\left(\frac{W}{4}\right) = {\sin^2(\rho/2) - \sin^2(\beta/2) \over \sin \alpha \cdot \sin \zeta},
        \label{WEq}
\end{equation}
where $\beta = (\zeta - \alpha)$ is the angle between the magnetic axis 
and the line of sight.
For RQ pulsars,  $\lvert\beta\rvert > \rho$ and $W$ is undefined. If the beam grazes 
Earth, $W$ may be so small that the
radio flux integrated over the narrow pulse is below a given radio 
telescope's sensitivity,
and if detected it is likely very faint: We call these borderline 
cases radio-faint (RF).

From the Pulsar Handbook, 
$\rho = A_1 f^{-q} + A_0$ or
$r_{em} = B_1 f^{-p} + B_0$, where $f$ is the radio frequency.
For positive values of the 
index $p$,
Eq.~\eqref{rhoEq} 
leads to beam widening at low radio frequencies.
This is called radius-to-frequency mapping (RFM).
The parameters $(B_0, B_1,p)$ have different values 
for different pulsars. 
For example, \citet{Thorsett91} 
studied seven multi-component pulsars that have 
been observed over a wide frequency range, 
finding the outer components to be separated by angles of
5-10$^{\circ}$ at 1400 MHz, 
but up to 30$^{\circ}$ at 150 MHz.
Similarly, \citet{Xilouris96} studied eight pulsars. Their Figure 2 shows 
the profile widths to approximately double between 1000 and 100 MHz.

While $W$ increasing at low frequency $f$ is common, this is not the case for all pulsars. 
\citet{Pilia16} determined the index $\delta$ for 100 pulsars, 
such that the observed width at 10\% of the pulse amplitude is $w_{10} \propto f^\delta$ (see their Figures 3 and 7).
For $\sim20$\% of the pulsars, they find $\delta > 0$, meaning that the pulse
narrows with decreasing frequency. 
However, they did indeed obtain  $\delta < 0$ for 80\% of the pulsars and 
found an average value of $\left< \delta \right> \sim -0.1$ (weighted mean over all values of $\delta$).
Thus, a significant fraction of their sample roughly agrees with the expectation from simple RFM.

Based on these observations, we assume RFM to hold. More specifically, we used
the parameterization of \citet[][Equations (9) and (10)]{Story07},
which is based on the model of \citet{Kijak97,Kijak98,Kijak03}.
In terms of the above parameterization, this is equivalent to $B_0=0$ and $p=0.26$.
Using $P\ = 100$ ms and $\dot{P}\ =\ 1\times10^{-15}$ s s$^{-1}$, 
this model
gives a beam width of 16.8$^{\circ}$ at 1400 MHz and 22.5$^{\circ}$ at 150 MHz.
In Section \ref{sec-results-beam-width-method}, we combine these values 
with geometrical constraints $(\alpha,\, \zeta)$ extracted from 
the gamma-ray profiles to predict radio detectability.
The comparison of these model results to pulsed emission discovered 
with FR606 (or the upper limits thereof) provides useful data for emission models.

\section{Methods} \label{sec-methods}

\subsection{Target selection} \label{sec-target-selection}

To a good approximation, the effective area of the FR606 antenna array is 
\begin{equation}
        A_{\rm eff} = A_\text{eff}^\text{max} \cos^2 z,
        \label{eq:Aeff}
\end{equation}
where $z$ is the zenith angle of the source and $ A_{\rm eff}^{max} =2048 \,\rm m^2$ \citep[][Appendix B]{vanHaarlem13}. 
For the telescope's latitude ($\lambda = 47.35^\circ$), 
a declination limit of 
$\delta> -10^\circ$ ensures $\cos^2 z > 0.28$ at pulsar culmination, 
so that the telescope can be used with a meaningful sensitivity for a few hours on a given target each day. 
We thus selected the 20 radio-quiet pulsars in 2PC with $\delta > -10^\circ$,
excluding Geminga 
(J0633+1746), which has already been extensively explored at low radio frequencies  \citep[][and references therein]{Maan15}.
In addition, we included seven
pulsars discovered 
after 2PC \citep{Pletsch13,Clark15,Clark17,Wu18}.
This leaves us with a total of 27 targets, which are 
listed in Table \ref{tab-pulsars} (column 1).

\begin{table*}[ht]
\tiny{
\begin{center}
\caption{Northern radio-quiet gamma-ray pulsars.}
\begin{tabular}{l|rrrr|rrcr|rrrrr}
pulsar &  $l$    &    $b$     & $P$   & $10^{-34}\dot E $  & $f_\text{ref}$ & $S_\text{ref}$ & Ref. & $S_{150}^{\text{extrapol}}$ & $N_\text{obs}$ & 
$T_\text{sky}$ & $t_\text{obs}^\text{eff}$ &$S_{150}^\text{min}$ & $s$ \\
    &($^\circ$) & ($^\circ$) & (ms)  & (erg s$^{-1}$)           & (MHz) & (mJy) &  & (mJy)& & (K) & (h) & (mJy) &\\
\hline \\
 J0002+6216$^{*}$  & 117.33 & -0.07 &  115.4  & 15.3  & 1400            & 0.022 {(detection)}     & 1,2,9 & $\sim$0.78    & 9 & 859 & 6.5 & $<$2.1 & $>-2.1$\\ 
 J0007+7303  & 119.66 & 10.46 &  315.9  & 44.8  & 820           & $<$0.012                                 & 3     & $<$0.18       & 9 & 619 & 5.4 & $<$1.7 \\ 
 J0106+4855$^{*}$  & 125.47 & -13.87 &  83.2  & 2.9   & 820             & 0.030 {(detection)}     & 4     & $\sim$0.45    &12 & 482 &10.4 & $<$1.3 & $>-2.2$\\ 
 J0357+3205  & 162.76 & -16.01 &  444.1 & 0.6   & 327           & $<$0.043                                 & 3     & $<$0.15       &12 & 374 & 8.9 & $<$0.8 \\ 
 J0359+5414  & 148.23 & 0.88 &  79.4    & 131.8 & 1400          & $<$0.015                                 & 1,2   & $<$0.53       &12 & 716 &10.5 & $<$1.7 \\ 
 J0554+3107  & 179.06 & 2.70 &  465.0   & 5.6   & 1400          & $<$0.066                              & 5     & $<$2.4  &11 & 483 & 7.6 & $<$1.2 \\ 
 J0622+3749  & 175.88 & 10.96 &  333.2  & 2.7   & 820           & $<$0.032                              & 4       & $<$0.48       & 7 & 457 & 5.6 & $<$1.4 \\ 
 J0631+0646$^{*}$  & 204.68 & -1.24 &  111.0  & 10.4  & 1400            & 0.018 {(detection)}     & 1,2,9 & $\sim$0.64& 7 & 641 & 2.0 & $<$3.2 & $>-2.3$ \\ 
 J0633+0632  & 205.09 & -0.93 &  297.4  & 11.9  & 1510          & $<$0.003                              & 3       & $<$0.12       & 9 & 641 & 2.0 & $<$2.3 \\ 
 J1836+5925  & 88.88 & 25.00 &  173.3   & 1.1   & 820           & $<$0.010                                 & 3     & $<$0.15       & 3 & 493 & 2.1 & $<$2.5 \\ 
 J1838$-$0537& 26.51 & 0.21 &  145.7    & 593.3 & 2000          & $<$0.009                              & 6       & $<$0.57       & 4 &4500 & 0.6 & $<$26  \\ 
 J1846+0919  & 40.69 & 5.34 &  225.6    & 3.4   & 1510          & $<$0.004                                 & 8     & $<$0.16       & 2 &1370 & 0.9 & $<$7.6 \\ 
 J1906+0722  & 41.22 & 0.03 &  111.5    & 102.2 & 1400          & $<$0.021                                 & 7     & $<$0.75       & 9 &2830 & 3.0 & $<$8.0 \\ 
 J1907+0602$^{*}$  & 40.18 & -0.89 &  106.6   & 282.4 & 1510            & 0.005 {(detection)}     & 10,3  & $\sim$0.20    &11 &2510 & 3.2 & $<$7.1 & $>-3.1$\\ 
 J1932+1916  & 54.66 & 0.08 &  208.2    & 40.7  & 1400          & $<$0.075                              & 5       & $<$2.7        & 8 &1410 & 4.1 & $<$2.9 \\ 
 J1954+2836  & 65.24 & 0.38 &  92.7     & 104.8 & 1510          & $<$0.004                              & 8       & $<$0.16       &11 &1150 & 8.2 & $<$2.1 \\ 
 J1957+5033  & 84.60 & 11.00 &  374.8   & 0.5   & 820           & $<$0.025                              & 8       & $<$0.38       &10 & 613 & 8.6 & $<$1.3 \\ 
 J1958+2846  & 65.88 & -0.35 &  290.4   & 34.2  & 1510          & $<$0.005                              & 3       & $<$0.20       &12 &1150 & 8.3 & $<$2.0 \\ 
 J2017+3625  & 74.51 & 0.39 &  166.7    & 1.2   & 1510          & $<$0.005                               & 1,2   & $<$0.20       &13 & 554 & 8.1 & $<$1.3 \\ 
 J2021+4026  & 78.23 & 2.09 &  265.3    & 11.4  & 2000          & $<$0.011                                 & 3     & $<$0.69       &13 &2830 &10.3 & $<$3.2 \\ 
 J2028+3332  & 73.36 & -3.01 &  176.7   & 3.5   & 1510          & $<$0.004                               & 4     & $<$0.16       & 7 & 921 & 5.2 & $<$1.9 \\ 
 J2030+4415  & 82.34 & 2.89 &  227.1    & 2.2   & 820           & $<$0.019                               & 4     & $<$0.29       &12 &1420 &10.1 & $<$2.2 \\ 
 J2032+4127$^{*}$  & 80.22 & 1.03 &  143.2    & 36.3  & 2000            & 0.05 {(detection)}      & 11, 3         & $\sim$3.2     &11 &1600 & 9.6 & $<$2.6 & $>-1.5$\\ 
 J2055+2539  & 70.69 & -12.52 &  319.6  & 0.5   & 327           & $<$0.085                                 & 8     & $<$0.30       & 9 & 509 & 5.3 & $<$1.3 \\ 
 J2111+4606  & 88.31 & -1.45 &  157.8   & 143.6 & 820           & $<$0.033                              & 4       & $<$0.50       & 9 & 886 & 7.2 & $<$1.9 \\ 
 J2139+4716  & 92.63 & -4.02 &  282.8   & 0.3   & 820           & $<$0.034                                 & 4             & $<$0.52       &11 & 730 & 9.9 & $<$1.1 \\ 
 J2238+5903  & 106.56 & 0.48 &  162.7   & 88.8  & 820           & $<$0.027                              & 3       & $<$0.41       &11 &1080 & 9.7 & $<$1.9 \\ 
\end{tabular}
\tablefoot{
Column 1: name of the pulsar (pulsars with know radio emission are denoted with $^{*}$).
Columns 2 and 3: Galactic coordinates for each pulsar.
Column 4: pulsar period ($P$).
Column 5: spindown luminosity $\dot E$. 
Columns 6 and 7: observing frequency and measured mean flux density 
(or upper limit) for the most constraining observation (assuming a spectral index of $-1.6$). 
Column 8: Reference (for columns 6 and 7). 
Column 9: flux density (or upper limit) from columns 6 and 7 extrapolated to an observing frequency of 150 MHz (using a spectral index of $-1.6$).
Column 10: the number of good observations (usually one hour) used in the final analysis (i.e., with a sufficiently low RFI level).
Column 11: sky temperature.
Column 12: the equivalent duration of the dataset under the assumption of nominal gain, as given by Eq. (\ref{eq-teff}).
Column 13: flux density upper limit at 150 MHz (in the band 110--190 MHz) as determined in this work.
Column 14: limit on the spectral index $s$ derived from the most constraining previous detection (columns 6 and 7) and our flux density limit (column 13).
\label{tab-pulsars}}
\tablebib{
(1) \citet{Clark17}, (2) \citet{Wu18}, (3) \citet{Ray11}, (4) \citet{Pletsch12_744}, (5) \citet{Pletsch13}, (6) \citet{Pletsch12_755}, (7) \citet{Clark15},
(8) \citet{SazParkinson10}, (9) J. Wu (personal communication), (10) \citet{Abdo10_711}, (11) \citet{Camilo09}.
}
\end{center}
}
\end{table*}

All of our targets have already been observed by radio telescopes, 
with observing frequencies between 34 and 2000 MHz (see Table 
\ref{tab-pulsars-observations} for the full list of observations). 
Table \ref{tab-pulsars} (columns 6--8) shows the flux density upper limits for 
the most constraining of the previous observations.
Detections rather 
than upper limits are denoted as such;
in those cases, 
we indicate the 
equivalent flux density, assuming $W/P=0.1$,
rather than the 
measured flux density, with the measured value for the
fractional pulse width $W/P$ 
(see Appendix \ref{sec-appendix}).
Column 9 gives the corresponding flux density limit at our observing 
frequency of 150 MHz for an assumed spectral index of $-1.6$. 

\subsection{Observations} \label{sec-observations}

The observations were carried out on the International LOFAR Station in Nan\c{c}ay, FR606.
LOFAR is fully described in \citet{Stappers11} and \citet{vanHaarlem13}.
LOFAR stations have two different frequency bands: We used the HBA band 
(i.e., 110--190 MHz, with a center frequency of 149.9 and a total bandwidth of 78.125 MHz), for which
the station consists  of 96 antenna tiles, each of which is made up of 16 dual-polarization antenna elements. 
The signals from individual HBA tiles were coherently summed, creating a digital telescope.

While a single LOFAR station as used here only has a limited effective area or telescope gain, it allows for very 
flexible scheduling, especially for long observations. The capability of this 
setup for pulsar science has already been demonstrated 
\citep{Rajwade16,Mereghetti16,Bondonneau17IAU,
Michilli18,Hermsen18,Donner19,Porayko19,Tiburzi19,Bondonneau20fr606}.   

Long observing sessions were split into individual observations typically lasting one hour.
Each target was observed between 7$\times$1h and 16$\times$1h, amounting to a total of 346h of telescope time 
(on average approximately 13h per target). 
To ensure phase coherence, the time span of our radio observations was at 
most 15 days for each pulsar.
The analysis of test data showed that daytime observations contained intense  radio frequency 
interference (RFI) and they did not provide data of a sufficiently good quality. 
To determine which fraction of the day allowed for high-quality observations, 
we observed a well-known pulsar (B0105+68, period $P\sim1.07$ s, 
DM$\sim61.06$ pc cm$^{-3}$) with the same setup and pipeline as for 
sources of interest. 
Even though the pipeline includes RFI cleaning, 
the pulsar was not detected in daytime observations, while it was clearly 
detected in all nighttime observations.
For this reason, all observations  of the sources of interest were taken at night.

Despite this precaution, a number of observations showed high RFI levels, 
and they had to be discarded subsequently (observations were discarded if RFI
led to false candidates at apparent $S/N>7$ even after RFI cleaning, see next section). 
In the end, only 255h of observations with low RFI levels were retained 
(apparent $S/N$$<7$), and 
the final analysis was based on between 2 and 13h per target (Table \ref{tab-pulsars}, 
column 10 gives the number of observing sessions lasting typically 1 hour; 
column 12, the equivalent duration of a zenith observations, $t_\text{obs}^\text{eff}$,  is explained
in Section \ref{sec-flux-limit}).

\subsection{Data analysis} \label{sec-data-analysis}

For the data analysis, we used the standard pulsar tools 
\texttt{tempo2} \citep{Hobbs06}, \texttt{DSPSR} \citep{vanStraten11},
 \texttt{PSRCHIVE} \citep{vanStraten12}, and \texttt{COASTGUARD}\footnote{https://github.com/plazar/coast\_guard/} \citep{Lazarus16}.
We folded the data at the period determined by the LAT rotation ephemeris, thus 
leaving the pulsar's dispersion measure (DM) as the only free parameter
during the data analysis.
In detail, we proceeded as described in the following.

				Shortly before the observation, an ephemeris was constructed using \textit{Fermi} LAT data.
				LAT radio-quiet pulsars are necessarily bright in gamma rays: 
                Faint gamma-ray sources are too poorly localized and their photon arrival times are too infrequent to allow
                for effective blind pulsation searches. The gamma-ray brightness of our pulsar sample made it easy to extend 
                their rotation ephemerides to cover the FR606 observation epochs 
                using the method detailed in \citet{Kerr15}. We started with the ephemerides available from the references 
                in Table \ref{tab-pulsars} for each pulsar. 
                These yield a constant gamma-ray phase to the end of the ephemeris validity, and well beyond for the more stable pulsars.
                From these, we made a template pulse profile using time intervals yielding profiles with a significance of $3\sigma$ 
                (typically a week to a month, depending on the flux). 
                Cross-correlating the profiles with the template determines a LAT time of arrival for each interval.
                With these times of arrival, \texttt{tempo2} \citep{Hobbs06} can then be used to improve the ephemeris. 
                If the gamma-ray pulse is lost beyond the end of ephemeris validity period, 
                it can generally be recovered after only a few iterations. 
                We thus obtained an ephemeris for each of the 27 pulsars 
                that is valid at the epoch of the FR606 observation.
                
				During the observation, the full data stream was split on four different data acquisition 
                machines (each receiving one quarter of the total bandwidth). 
                The \texttt{LuMP} software\footnote{https://github.com/AHorneffer/lump-lofar-und-mpifr-pulsare}
                was used to recorded the data.
                
				After the observation, the four datasets (one per data acquisition machine) 
                were each folded at the initial \textit{Fermi} LAT period using \texttt{DSPSR} \citep{vanStraten11}
                and channelized into 2440 channels of $\sim$10 kHz. 
                The number of channels was chosen as a compromise between the available 
                computing resources and the expected loss of 
                sensitivity due to DM smearing (the associated factor $\beta_\text{$\delta$DM}$ 
                is quantified in Section \ref{sec-flux-limit}).
                Subsequently, the borders of the bands were removed (low sensitivity of the antennas near
                the edge of their bandpass), 
                so that in total 8000 useful frequency channels were kept (the frequency range of 110--190 MHz).
                
				Then, the data were cleaned of RFI using \texttt{COASTGUARD} \citep{Lazarus16}. 
                Typically, $\sim 3$\% of the data retained in the final step were flagged (with a maximum flagged fraction of $8.4$\%).
                
				Subsequently, the data were rebinned to subintegrations of 300 s in order to reduce the 
                data volume and speed up the post-processing.
				In the next step, the data from the four acquisition machines were combined.
				
				A few months after the observation, an improved ephemeris based on \textit{Fermi} LAT data was constructed
                (based on the extended \textit{Fermi} LAT time series).
                The datasets (the 300 s subintegrations) were refolded with this improved ephemeris.
                
				Using \texttt{pdmp}, we searched each individual observation session ($\sim$1 hour) for a dispersed,                         
                periodic radio signal at the \textit{Fermi} LAT period. 
                This search was performed for all pulsars, including 
                those already detected in radio: 
                The DM precision of the high frequency observations might not 
                be good enough (the effect of a small DM error is 100 times 
                stronger at 150 MHz than at 
                1500 MHz); also, 
                the precise DM value could have changed since the detection.
                We used 3980 trial DM 
                values between 2 and 400 pc/cm$^{3}$ in steps of 0.1 pc/cm$^{3}$ (incoherent dedispersion). 
                For 19 of our 27 lines of sight, the Galactic electron density model 
                NE2001 \citep{Cordes02} predicts a maximum DM of less than 400 pc/cm$^{3}$.
                For the eight remaining lines of sight (all close to the Galactic plane), the corresponding estimated scattering time at 150 MHz 
                                        (assuming a scatter broadening spectral index of $-4$)
                already exceeds the pulsar's period, rendering an extension to even higher DM values useless.
                The result is similar for the YMW16 model \citep{Yao17}: For 17 lines of sight, the maximum DM is below 400 pc/cm$^{3}$, 
                and for most of the remaining ten lines of sight, the estimated scattering time at 150 MHz exceeds the pulsar's period.
                The only exceptions are 
                J1958+2846, J2021+4026, and J2030+4415, for which a maximum DM of $\sim$500 would have been a slightly better choice.
                
				Manual inspection showed that all observation sessions for which \texttt{pdmp} produced a high signal-to-noise
				ratio (S/N) 
                were in fact corrupted by remaining RFI and showed high values for a wide range of DM values.
                For this reason, all observation sessions with a value of apparent $S/N \ge 7$ were discarded.
				The remaining observation sessions of each target (between two and 13 sessions depending on the target)
                were added together using \texttt{psradd}.
                
				For the combined files (one per target), we searched again for dispersed, periodic signals using 
                \texttt{pdmp}. All candidate detections with a value of $S/N \ge 5$ (a total of 1657 candidates) were inspected manually.

This procedure was tested and validated on a well-known pulsar 
(B0105+68, period $P\sim1.07$ s, $DM\sim61.06$ pc cm$^{-3}$).
We then processed the observations of the 27 pulsars of Table \ref{tab-pulsars} in this way; 
no pulsed radio signal was detected. To quantify our nondetections, 
we calculated the flux density upper limits, 
taking the elevation of the pulsar during each observing session into account (see following section).

\subsection{Flux density upper limits} \label{sec-flux-limit}

In none of our observations was a pulsar radio signal detected. 
Upper limits for the pulse-average flux were calculated following the Pulsar Handbook:
\begin{equation}
        S_\text{min} = \frac{
        S/N_\text{min} \left(T_\text{rec}+T_\text{sky}\right)}
        {\beta_\text{$\delta$DM} G_\text{nom} \sqrt{n_p t_\text{obs}^\text{eff} 
        \Delta F_\text{eff}}} \sqrt{\frac{W}{P-W}}\,,
\end{equation}
where a value of $S/N_\text{min}=5$ was assumed as a threshold in the S/N 
required for a detection, $T_\text{rec}=422\,\text{K}$ is the receiver temperature
\citep[as derived from measurements by][]{Wijnholds11ieee}.
The sky temperature
$T_\text{sky}$ depends on the sky position of the observed target; it is 
taken from the sky map at 408 MHz by 
\citet{Haslam82} and was scaled to our frequencies using $f^{-2.55}$ \citep{Lawson87}.
The values for $T_\text{sky}$ are given in Table \ref{tab-pulsars} (column 11). 
Furthermore, $\beta_\text{$\delta$DM}$ is the loss of sensitivity due to imperfect DM gridding (see below), 
$G_\text{nom}$ is the 
nominal telescope gain for a pointing near zenith, 
$n_p$ is the number of polarizations (in all of our observations, $n_p=2$), 
$t_\text{obs}^\text{eff}$ is the effective observing time for each pulsar (see below), 
and $\Delta F_\text{eff}$ is the effective frequency bandwidth of each observation after RFI cleaning. 
Finally, $P$ is the pulsar period and $W$ is the width of the pulse; for all pulsars, 
we assumed $W/P=0.1$.

With our grid of trial DMs, the DM error $\delta$DM of the best trial is at most 0.05 pm/cm$^3$. This leads to a slight loss in the S/N and thus in sensitivity.
Using 
\citet[][eq. (12--13)]{Cordes03ApJ}, a center frequency of 149.9 MHz, a total bandwith of 78.125 MHz, and $W/P=0.1$, 
we find values in the range from $0.75\le\beta_\text{$\delta$DM}\le0.99$ for our sample, depending on the pulsar's period. 
These values were taken into account for the calculation of the flux density limits in Table \ref{tab-pulsars}.

For the nominal gain, 
we used 
$G_\text{nom}={A_{\rm eff}^\text{max}}/{(2 k_B)}=0.74 \text{K}/\text{Jy}$, 
which is equivalent to the effective area of $A_\text{eff}^\text{max}=2048 $m$^2$
of the international LOFAR station 
FR606 
for a pointing near zenith at 
an observing frequency of 150 MHz
\citep[][Appendix B]{vanHaarlem13}. 
As the elevation of the pulsar varies during 
an observation, the projected area of the antenna array varies over time, cf.~eq.~(\ref{eq:Aeff}). 
Assuming the effective area is approximately constant over an observation of 
approximately one hour, 
the effective observing time after 
$N_{\text{obs}}$ observations
(with an effective area $A_{\text{eff},i}$ for the $i^{th}$ observation) is given by 
\begin{equation}
        t_\text{obs}^\text{eff} = \sum_{i=1}^{N_{\text{obs}}} t_i \left(\frac{A_{\text{eff},i}}{A_\text{eff}^\text{max}}\right)^2. \label{eq-teff}
\end{equation}
It is important to note that this implies that observations at a low elevation (and thus with $A_{\text{eff},i} \ll A_\text{eff}^\text{max}$) 
contribute only marginally to the combined observation.
The values of $t_\text{obs}^\text{eff}$ are given in Table \ref{tab-pulsars} (column 12). 

The values of $A_{\text{eff},i}$ and $T_\text{sky}$ were calculated using the LOFAR 
calibration tool \texttt{lofar\_fluxcal.py} \citep{Kondratiev15}. 
The software makes use of the Hamaker beam model \citep[][and references therein]{Hamaker06} 
and the \texttt{mscorpol}\footnote{https://github.com/2baOrNot2ba/mscorpol} 
package 
to calculate the Jones matrices for a given frequency 
and sky coordinates. 

With this, the upper limits on the flux density, $S_\text{min}$, were calculated, and 
the values are given in Table \ref{tab-pulsars} (column 13). 
The implications are discussed in Section \ref{sec-results}.

\section{Results}
\label{sec-results}


\begin{figure*}
        \includegraphics[width=0.9\linewidth]{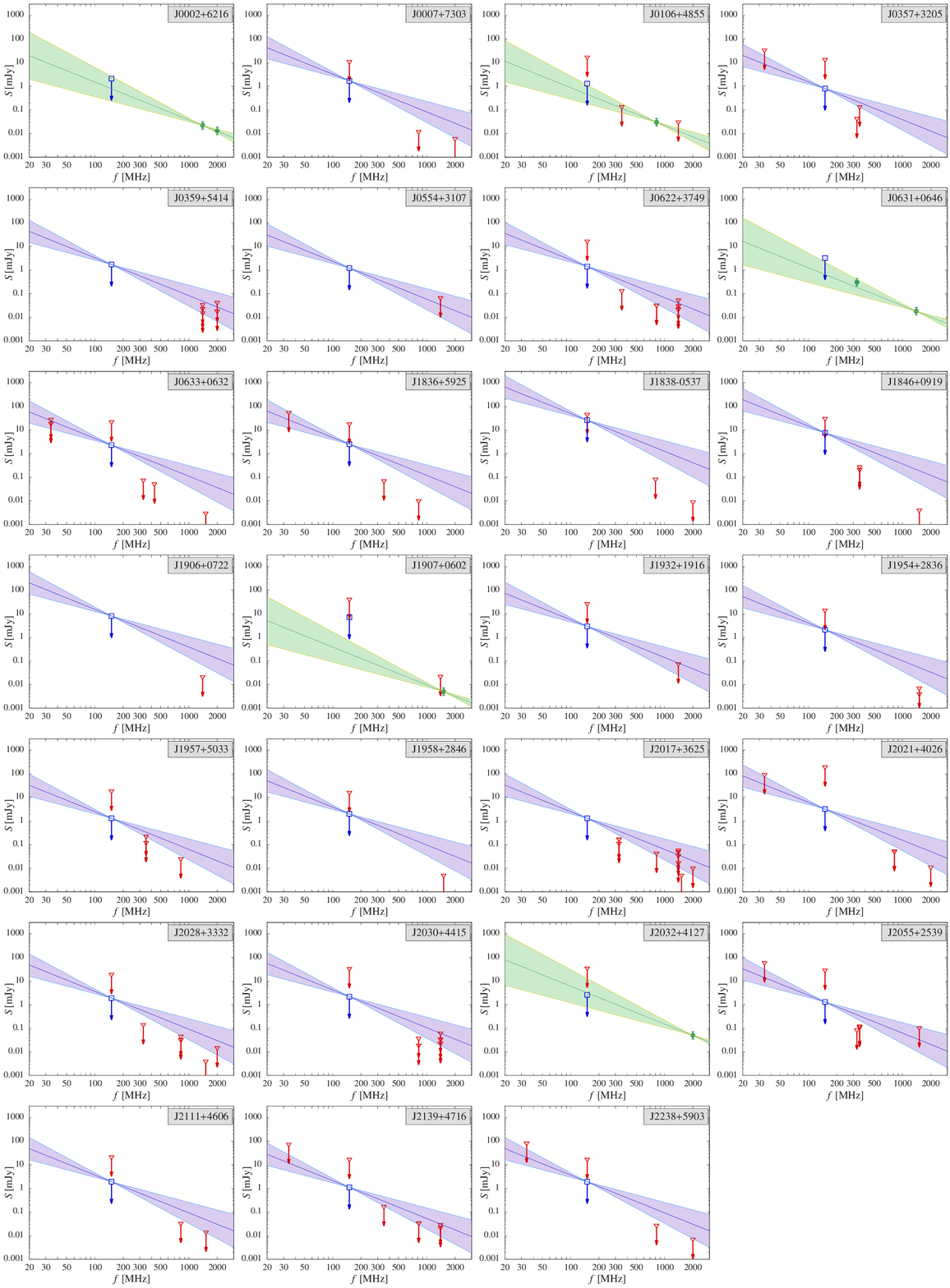}
        \caption{Flux density upper limits for 27 radio-quiet gamma-ray pulsars, based on observations with LOFAR/FR606.
        Empty red triangles (with arrows): Flux density limits from previous observations. 
        Filled green circles: Flux density measurements from previous observations (a typical error bar of 30\% is indicated). 
        Empty blue squares (with arrows): Flux density limits from this study. 
        Dark blue line: Equivalent flux density limits for our observation (assuming a spectral index of $-1.6$).
        Shaded area between the light blue lines: Equivalent flux density limits for our observation for a spectral index of $-1.6\pm0.54$.
        Dark green line: Same as the dark blue line, but with respect to the most constraining detection.
        Shaded area between the light green lines: Same as the shaded area between the light blue lines, but with respect to the most constraining detection.
        \label{fig-upperlimits}} 
\end{figure*}

No radio emission was detected for any of our targets.
At our observing frequency of 150 MHz, we obtained flux density upper limits between 0.8 mJy and 26 mJy
(the values are given in column 13 of Table \ref{tab-pulsars}).
The large differences in the upper limits for different targets result from four effects:
(a) The number of 1h observations was different for different targets;
(b) during some nights, the RFI conditions were worse than during others, forcing us to remove a larger number of observations from processing;
(c) the low elevation of some sources led to a low effective observing time $t_\text{obs}^\text{eff}$ despite a large number of observations $N_\text{obs}$; and
(d) depending on the direction, the background sky temperature $T_\text{sky}$ varied by more than one order of magnitude.
In particular, we note that
our flux density limits are less constraining for 
J1838$-$0537 (26 mJy),  J1846+0919 (7.6 mJy), J1906+0722 (8.0 mJy), and  J1907+0602 (7.1 mJy) 
than for our other targets. 
This is related to their proximity to the Galactic plane (Galactic coordinates are given in 
Table \ref{tab-pulsars}, columns 2 and 3), 
which causes a high sky temperature ($T_\text{sky}$, Table \ref{tab-pulsars} column 11), 
and at the same time this reduces the effective area of the telescope because of their low declination 
(and thus leads to a small $t^\text{eff}_\text{obs}$, column 12).
We obtained our most constraining flux density limit of 0.8 mJy for J0357+3205, 
where the sky temperature is particularly low 
(far off the Galactic plane) and 
for which RFI conditions were good, so that we have a large number of usable observations.

Table \ref{tab-pulsars} shows the most constraining previous observation for each pulsar,
assuming a spectral index of $-1.6$. The observing frequency and
the upper limit for the flux density are given in columns 6 and 7,
and the equivalent flux density at 150 MHz is shown in column 9.

The same data are displayed in Figure \ref{fig-upperlimits}\footnote{
For detected pulsars, we want to compare the upper limits to measured values. 
For this reason, we used the nominal flux density values, assuming $W/P=0.1$, 
rather than the measured flux density (with the measured value for the
fractional pulse width $W/P$),
see Appendix \ref{sec-appendix}.}, 
in which we compare 
our flux density upper limits (empty blue squares)
to upper limits (empty red triangles) and detections (filled green circles,
with a typical error bar of 30\%)
obtained by previous observation campaigns 
(all values are summarized in Appendix \ref{sec-appendix} and 
Table \ref{tab-pulsars-observations}).
As can be seen in  Figure \ref{fig-upperlimits}, our nondetections are 
compatible with the nondetection of those 21 of our targets that have been
studied with imaging observations at 147.5 MHz using GMRT \citep{Frail16}; 
our observations do, however, provide more stringent upper limits (by approximately one order of magnitude).
The remaining six pulsars have never before been observed in this frequency range.

To test whether our upper limits are compatible with previous observations at different frequencies,
we assume a spectral index of $-1.6\pm0.54$
(see Section \ref{sec-spectral-index}).
To compare this to previous nondetections, the dark blue lines  
in Figure \ref{fig-upperlimits} 
represent
flux density limits equivalent to our observation,
assuming an average spectral index of $-1.6$, and
the shaded areas between the light blue lines correspond to a spectral index 
of $-1.6\pm0.54$ (see Section \ref{sec-spectral-index}).
Figure \ref{fig-upperlimits} shows that our upper limits are compatible with 
all previous upper limits.
For J0554+3107, our upper limit is more constraining 
than the previous observation by a factor $\sim2$.

Five of the pulsars we observed have been detected at higher radio frequencies,
but they remain undetected in our observations. 
In Table \ref{tab-pulsars}, these pulsars are denoted with $^{*}$ in column 1, 
and limits for the spectral index $s$ are given in column 14.
For these pulsars,
the dark green lines  
in Figure \ref{fig-upperlimits} 
represent
flux density limits equivalent to the most constraining detection, 
assuming an average spectral index of $-1.6$, and
the shaded areas between the light green lines correspond again to a spectral index 
of $-1.6\pm0.54$ (see Section \ref{sec-spectral-index}).

For J0002+6216, radio emission has been reported at frequencies of 1400 and 2000 MHz
\citep{Clark17,Wu18}.
When combined, these two observations hint at a spectral index shallower than average 
($-1.2$). However, as both observing frequencies are not widely separated, 
the spectral index is not very well constrained. 
Our nondetection is compatible with both previous observations and constrains
the spectral index to values $>-2.1$.

The pulsar J0106+4855 has been detected at 820 MHz \citep{Pletsch12_744}.
With our setup, a nondetection is expected as long as the spectral index
is $>-2.2$.

J0631+0636 has been detected at 327, 1400, and 1510 MHz.
The two available flux measurements 
hint at a spectral index that is steeper than average ($-1.9$). 
This spectral index is still compatible with our nondetection, which constrains the spectral index to values 
$>-2.3$.

For J1907+0602, very faint radio emission has been reported at 1510 MHz \citep{Abdo10_711,Ray11}.
This is consistent with our nondetection and constrains the spectral index to values $>-3.1$.

Finally, for J2032+4127, faint radio emission has been reported at 2000 MHz \citep{Camilo09,Ray11}. 
Extrapolating their flux measurement to our observing frequency with an assumed spectral index of $-1.6$ gives a flux value 
20\% higher than the upper limit derived from our observations.
Considering typical errors on flux density measurements and the variability of pulsars, 
our nondetection is compatible with the detection at 2000 MHz.
It can also be explained by either 
variations in the pulsar's emission (either intrinsic or due to scintillation on a timescale larger than the duration of the observation), 
or a spectral break or 
turnover 
in the 100--200 MHz range.
Alternatively, it could be undetectable because of a flatter spectral index than the average. 
Indeed, for a spectral index shallower than $-1.5$, which is compatible with the range of spectral indices discussed above, 
the extrapolated flux is compatible with our nondetection.

In all cases, our observations provide the most constraining upper limits at 
150 MHz for the set of 27 pulsars we observed.
Based on the comparison above, J0631+0636 and 
J2032+4127 would be the most interesting targets for potential reobservations at 150 MHz.

\section{Beaming, revisited} \label{sec-results-beam-width}

\subsection{Simple expectations} \label{sec-results-beam-width-motivation}

In Section \ref{sec-beam-widening}, for simple RFM, we have provided a beam  
width of $\rho_{1400}=16.8^{\circ}$ at 1400 MHz, 
and $\rho_{150}=22.5^{\circ}$ for the HBA range (150 MHz).
The angle $\zeta$ can have values in the range from 
$0^{\circ}\le \zeta \le90^{\circ}$.

For any given 
value of $\alpha$, the pulsar is geometrically visible (potentially RL) if 
the line of sight is not too far from the magnetic axis, 
$(\alpha - \rho) < \zeta < (\alpha + \rho)$.
Uncertainties in $\alpha$ and $\zeta$ ($\Delta\alpha$ and $\Delta\zeta$) 
are accounted for, such that the 
pulsar is only classified as RL if this condition is satisfied for all 
values of $\alpha\pm\Delta\alpha$ and $\zeta\pm\Delta\zeta$.
If 
$\zeta < \alpha - \rho$ or $\alpha + \rho< \zeta$
(again, accounting for the uncertainties in 
$\alpha$ and $\zeta$),
the pulsar
 is considered as RQ. 
In all other cases (the pulsar could be either RL or RQ depending on
the precise values of $\alpha$ and $\zeta$ within the uncertainties),
the pulsar is considered RF. 

Figure \ref{fig:geometry} shows the ($\alpha$, $\zeta$) plane.
The values shown for the pulsars are explained in the next section.
A few of these pulsars should be geometrically visible. This is 
expected, for example, for the pulsar with ($\alpha=85^\circ$, $\zeta=84^\circ$) 
in Figure \ref{fig:geometry-og}.
As the beam is wider at a low frequency ($\rho_{150}>\rho_{1400}$), 
a pulsar can be RQ at 1400 MHz, but RL at 150 MHz.
If we assume a fully uniform distribution of $\alpha$ and $\zeta$
and no uncertainties ($\Delta\alpha=\Delta\zeta=0$),
approximately $20$\% of the pulsars that are RQ at 1400 MHz could be RL at 150 MHz.
For our sample of 27 pulsars, this could mean on the order of five detections 
at 150 MHz with FR606. 

Of course, a uniform distribution of $\alpha$ and $\zeta$ is not expected,
and the uncertainties of $\alpha$ and $\zeta$ have to be taken into account.
In the following, we estimate the values of $\alpha$ and $\zeta$ for the
27 pulsars of our sample, based on their gamma-ray profiles, 
and we attempt to improve our estimate of how many
pulsars may have been detectable.

\subsection{Pulsar geometry from gamma-ray profiles} \label{sec-results-beam-width-method}
Gamma-ray beams are very narrow in neutron star longitude due to concentration of the gamma-radiating electrons
and positrons along ``caustically'' focused magnetic field lines; however, they are very broad in latitude, being
brightest near the neutron star equator, and fading toward the poles. 
Beam shapes depend on several parameters: The open field line configuration varies with $\alpha$ and $r_\text{LC}$, and
electron acceleration gap sizes depend on magnetic field strength. Gamma-ray emission models (see below)
differ in how they exploit these parameters, but all have in common that  ``in fine'' an observed profile depends
on $\alpha$ and $\zeta$. In the following we use the observed gamma-ray profiles 
to constrain the pulsar geometry (in terms of $\alpha$ and $\zeta$),
and thus refine our prediction of how many, and which, of our pulsars
may be radio-detectable at low frequency.

We first made weighted gamma-ray pulse profiles for our sample, integrated above 100 MeV.
Phases were calculated using \texttt{tempo2} 
and rotation ephemerides 
were derived from LAT data for use in 3PC
using the methods of \citet{Kerr15}.
Weights, that is the probability that a given photon comes from the pulsar and not from a background source, 
were calculated using the methods of \citet{Kerr11}.
In all cases, the lightcurves were compatible with those previously published, such as in 2PC.

The weight calculation requires a spectral and spatial analysis of the region 
surrounding the pulsar.  
For our analysis, we used the results of the FL8Y source 
list\footnote{See \url{https://fermi.gsfc.nasa.gov/ssc/data/access/lat/fl8y/}.} 
applied to Pass 8 (P8R2) LAT data corresponding to events within 
3$^{\circ}$ of each pulsar spanning 
2008-08-04 to 2016-08-04.  
Our timing solution for 
J1932+1916 folds poorly beyond 
2013-07-04. 
and so we used just under 5 years of data for that pulsar. 
Our LAT data sets include events belonging to the \texttt{SOURCE} 
class as defined under the \texttt{P8R2\_SOURCE\_V6} instrument response 
functions\footnote{See \url{https://www.slac.stanford.edu/exp/glast/groups/canda/lat_Performance.htm}.}, 
with energies between 0.1 and 100 GeV and zenith angles $\leq90^{\circ}$.

We then compared the gamma-ray profiles with the predictions from the outer gap (OG) model
and the 
two-pole caustic (TPC) model, generated over a broad parameter space. 
Specifically, we used the same simulations and pulse-profile fitting techniques as 
\citet{Wu18} \citep[for more details on the simulation and fitting see][]{Johnson14}. References to the
OG and TPC models are given there.
For 
J0002+6216, J0106+4855, J0631+0646, J1907+0602, and J2032+4127, 
we jointly fit the gamma-ray and radio profiles, using the 1400, 820, 1400, 1400, and 2000 MHz pulse profiles, respectively.  
As in \citet{Wu18}, we could not produce acceptable fits for J0631+0646.
Compared to \citet{Wu18}, our analysis includes more
gamma-ray photons, leading to updated
results, especially for J0002+6216 and J2017+3625.

Table \ref{tab-pulsars-angles} lists the resulting $\alpha$ and $\zeta$ angles, with
columns 2 and 3 for the OG model, and columns 4 and 5 for the TPC model. 
These results are also shown in Figure \ref{fig:geometry}.
Most of the fits are limited by systematic uncertainties 
of $\sim10^{\circ}$ \citep[see][for more details]{Johnson14,Pierbattista15}.

From the best-fit geometries and using the radio cone model 
of \citet{Story07}, we predicted the radio detectability for each pulsar
at 1400 MHz (Table \ref{tab-pulsars-radioflag}, columns 2 and 4 for models TPC and OG) and 150 MHz 
(Table \ref{tab-pulsars-radioflag}, columns 3 and 5 for models TPC and OG, respectively). 
RL means that our line of sight should intersect a bright part of the cone.
RF suggests that our line of sight skims the cone edge, 
and RQ means that the cone completely misses Earth.
Aligned and nearly aligned rotators are considered as RF.

The characteristic width of the simulated radio beam at 0.1\% of the peak intensity 
is given by Equations (9) and (10) of \citet{Story07}. 
For our simulations, we used $P\ = 100$ ms and $\dot{P}\ =\ 1\times10^{-15}$ s s$^{-1}$, 
which gives a width of 16.8$^{\circ}$ at 1400 MHz and 22.5$^{\circ}$ at 150 MHz.

Column 6 in Table \ref{tab-pulsars-radioflag} reports 
$(-\ln(\rm{likelihood})_{\rm TPC}\ -\ (-\ln(\rm{likelihood})_{\rm OG})$.  
When this value is positive, the OG model describes the data better than the TPC model, and vice versa.  
Based on their experience fitting many pulsars, \citet{Johnson14} found that
an absolute value of $\geq$15 was needed in
log likelihood difference to determine that one model 
was significantly favored over another.  
Based on this, the last column gives the preferred model for each pulsar
(indicated in parentheses if the log likelihood difference is $<$15).
Similar to the conclusions drawn 
by \citet{Johnson14}, \citet{Pierbattista15}, and others, we did not find 
that one model predominantly describes gamma-ray pulse profiles better than another. 

\subsection{Geometry results and discussion} \label{sec-results-beam-width-discussion}

\begin{figure}
\begin{center}
        \subfigure[TPC model]
        {\includegraphics[angle=0]
        {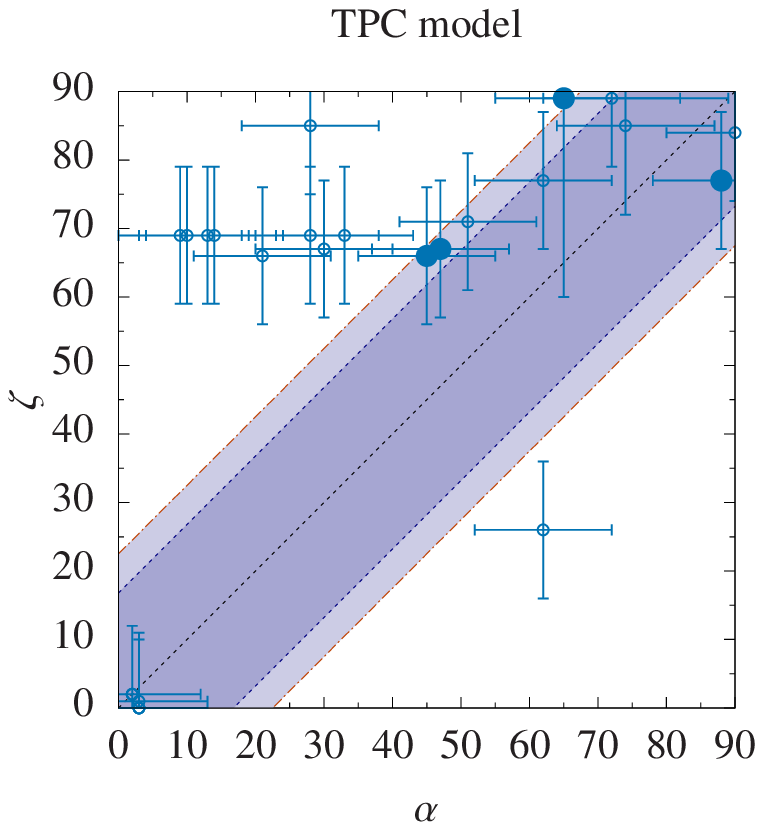}
        \label{fig:geometry-tpc}}
        \subfigure[OG model]
        {\includegraphics[angle=0]
        {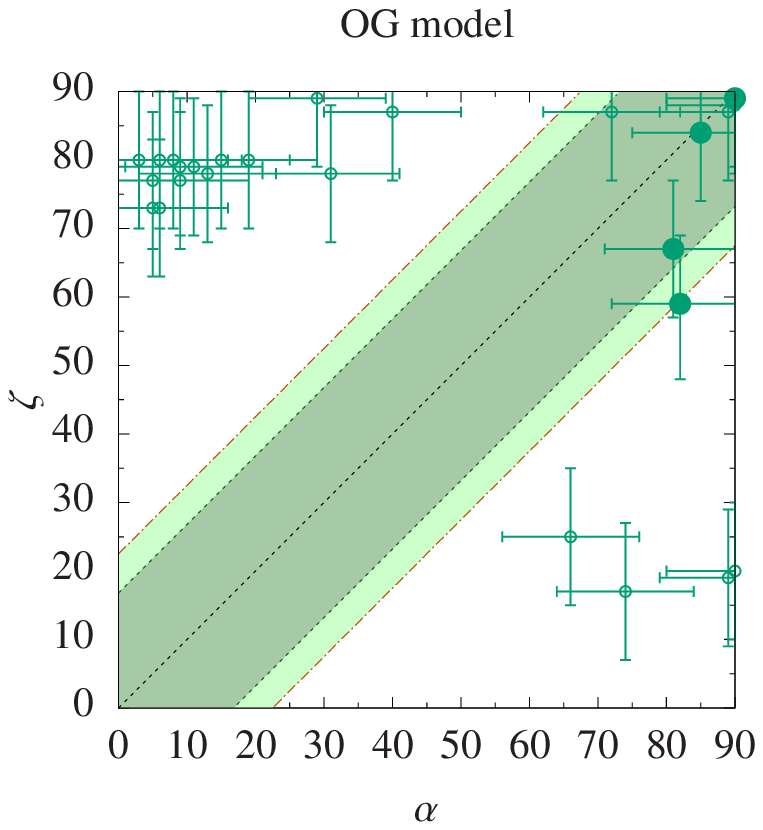}
        \label{fig:geometry-og}}
\caption{Pulsar viewing geometry, expressed by the angles $\alpha$ and $\zeta$, 
obtained by a fit to the gamma-ray data with the TPC model (panel a) and with the OG model (panel b). 
Pulsars on the dashed diagonal line ($\alpha=\zeta$) beam their radio emission to Earth regardless of their beam 
opening angle. Radio emission from pulsars between the two colored dashed lines is only detectable if their respective 
beam width $\rho\le16.8^{\circ}$, whereas pulsars between the two dash-dotted lines are visible 
if  $\rho\le22.5^{\circ}$.
Pulsars detected in radio are shown with large filled circles; pulsars without detected radio emission are shown with small empty circles.
\label{fig:geometry}}
\end{center}
\end{figure}

\begin{table}[ht]
\tiny{
\begin{center}
\caption{Pulsar geometry as derived from fitting the gamma-ray light curves.}
\begin{tabular}{l|rr|rr}
Pulsar & $\alpha_\text{TPC}$ & $\zeta_\text{TPC}$ & 
         $\alpha_\text{OG}$  & $\zeta_\text{OG}$  \\
       & (deg)               & (deg)              &
         (deg)               & (deg)              \\
\hline \\
J0002+6216$^{*}$ & 47 & 67 & 81 & 67 \\
J0007+7303       & 2  & 2  & 9  & 79 \\
J0106+4855$^{*}$ & 65$^{+24}_{-10}$ & 89$^{+10}_{-29}$ & 90 & 89 \\
J0357+3205       & 3 & 0 & 3  & 80  \\
J0359+5414       & 3 & 0 & 89 & 19  \\
J0554+3107       & 3 & 1 & 9  & 77  \\
J0622+3749       & 9 & 69 & 6 & 80 \\
J0631+0646$^{*}$ & \nodata & \nodata & \nodata & \nodata \\
J0633+0632       & 74$^{+13}_{-10}$ & 85$^{+10}_{-13}$ & 90 & 88 \\
J1836+5925       & 28 & 85 & 5 & 77  \\
J1838$-$0537     & 10 & 69 & 13 & 78 \\
J1846+0919       & 2  & 2  & 5  & 73 \\
J1906+0722       & 30 & 67 & 74 & 17 \\
J1907+0602$^{*}$ & 45 & 66 & 82 & 59$^{+10}_{-11}$ \\
J1932+1916       & 3  & 0  & 90 & 20 \\
J1954+2836       & 62 & 77 & 8  & 80 \\
J1957+5033       & 2  & 2  & 6  & 73 \\
J1958+2846       & 33 & 69 & 31 & 78 \\
J2017+3625       & 51 & 71 & 15 & 80 \\
J2021+4026       & 62 & 26 & 29 & 89 \\
J2028+3332       & 28 & 69 & 40 & 87 \\
J2030+4415       & 72 & 89 & 72 & 87 \\
J2032+4127$^{*}$ & 88 & 77 & 85 & 84 \\
J2055+2539       & 14 & 69 & 66 & 25 \\
J2111+4606       & 13 & 69 & 11 & 79 \\
J2139+4716       & 21 & 66 & 19 & 80 \\
J2238+5903       & 90 & 84 & 89 & 87 \\
\end{tabular}
\tablefoot{
Column 1: name of the pulsar (pulsars with know radio emission are denoted with $^{*}$).
Columns 2 and 3: best-fit values for the angles $\alpha$ and $\zeta$ for the TPC model.
Columns 4 and 5: same as columns 2 and 3, but for the OG model.
Unless otherwise stated, the error was considered to be $\pm{10}$.
\label{tab-pulsars-angles}}
\end{center}
}
\end{table}

\begin{table*}[ht]
\tiny{
\begin{center}
\caption{Radio-loud and radio-quiet pulsars.}
\begin{tabular}{l|rr|rr|cc}
Pulsar & Flag$_\text{TPC}^\text{150 MHz}$ & Flag$_\text{TPC}^\text{1400 MHz}$ & 
         Flag$_\text{OG}^\text{150 MHz}$  & Flag$_\text{OG}^\text{1400 MHz}$
           & $\Delta$(-ln(likelihood)) & preferred  \\
                &               &                       &       
                                                &           &                       
         (TPC-OG)       & model\\
\hline \\
J0002+6216$^{*}$ & RF & RF & \doublebox{RF} & \doublebox{RF} & 54.3  & OG\\
J0007+7303       & RF & RF & \doublebox{RQ} & \doublebox{RQ} & 548.1 & OG\\
J0106+4855$^{*}$ & RF & RF & \fbox{RL} & \fbox{RL} & 7.9   & (OG)\\
J0357+3205       & RF & RF & \doublebox{RQ} & \doublebox{RQ} & 170.5 & OG\\
J0359+5414       & RF & RF & \fbox{RQ} & \fbox{RQ} & 3.28 & (OG)\\
J0554+3107       & RF & RF & \fbox{RQ} & \fbox{RQ} & 5.5368 & (OG)\\
J0622+3749       & \fbox{RQ} & \fbox{RQ} & RQ & RQ & $-$10.2 & (TPC)\\
J0631+0646$^{*}$ & \nodata & \nodata & \nodata & \nodata & \nodata & \nodata\\
J0633+0632       & \doublebox{RL} & \doublebox{RF} & RL & RL & $-$111.6 & TPC \\
J1836+5925       & \doublebox{RQ} & \doublebox{RQ} & RQ & RQ & $-$2889.0 & TPC \\
J1838$-$0537     & RQ & RQ & \doublebox{RQ} & \doublebox{RQ} & 31.7 & OG\\
J1846+0919       & RF & RF & \fbox{RQ} & \fbox{RQ} & 8.7 & (OG) \\
J1906+0722       & \doublebox{RQ} & \doublebox{RQ} & RQ & RQ & $-$22.0 & TPC\\
J1907+0602$^{*}$ & RF & RF & \doublebox{RF} & \doublebox{RF} & 316.5 & OG\\
J1932+1916       & RF & RF & \doublebox{RQ} & \doublebox{RQ} & 19.1 & OG\\
J1954+2836       & \doublebox{RF} & \doublebox{RF} & RQ & RQ & $-$32.2 & TPC\\
J1957+5033       & RF & RF & \doublebox{RQ} & \doublebox{RQ} & 35.5 & OG\\
J1958+2846       & RQ & RQ & \doublebox{RQ} & \doublebox{RQ} & 336.1 & OG\\
J2017+3625       & \doublebox{RF} & \doublebox{RF} & RQ & RQ & -46.1 & TPC \\
J2021+4026       & \doublebox{RQ} & \doublebox{RQ} & RQ & RQ & $-$111.6 & TPC\\
J2028+3332       & RQ & RQ & \doublebox{RQ} & \doublebox{RQ} & 235.9 & OG\\
J2030+4415       & \doublebox{RF} & \doublebox{RF} & RF & RF & $-$100.9 & TPC\\
J2032+4127$^{*}$ & \doublebox{RL} & \doublebox{RF} & RL & RL & $-$470.6 & TPC\\
J2055+2539       & RQ & RQ & \doublebox{RQ} & \doublebox{RQ} & 82.5 & OG\\
J2111+4606       & RQ & RQ & \doublebox{RQ} & \doublebox{RQ} & 42.6 & OG\\
J2139+4716       & \doublebox{RQ} & \doublebox{RQ} & RQ & RQ & $-$19.8 & TPC \\
J2238+5903       & \doublebox{RL} & \doublebox{RL} & RL & RL & $-$295.8 & TPC\\
\end{tabular}
\tablefoot{
Column 1: name of the pulsar (pulsars with know radio emission are denoted with $^{*}$).
Columns 2 and 3: prediction regarding observability of the pulsar at 150 and 400 MHz for the TPC model: RL=radio-loud, RF=radio-faint, RQ=radio-quiet.
The flag is framed by a box if TPC is the preferred model for this pulsar (double box if the absolute value of column 6 is $>15$, see text for details). 
Columns 4 and 5: same as columns 2 and 3, but for the OG model.
Column 6: difference in likelihood for TPC and OG model (see text for details).
Column 7: preferred model (based on column 6). 
The model is indicated in parentheses if the absolute value of column 6 is $<$15.
\label{tab-pulsars-radioflag}}
\end{center}
}
\end{table*}

Table \ref{tab-pulsars-angles} shows 
our results for the geometric angles $\zeta$ and $\alpha$ 
for the models TPC and OG.
The implications for (geometrical) 
detectability of radio emission at 150 MHz and 1400 MHz
are shown in Table \ref{tab-pulsars-radioflag}.
The results are also shown in Figure \ref{fig:geometry}, where
pulsars detected in radio are shown with large filled circles
and pulsars without detected radio emission are shown with small empty circles.

In Table \ref{tab-pulsars-radioflag}, 
pulsars with detected radio emission are denoted with a $^{*}$ in column 1.
It can be seen that all of these
are labeled as either RF
or RL. 
Equivalently, all detected pulsars fall within or close to the shaded region in Figure 
\ref{fig:geometry}.

Only one of the pulsars labeled as RL at 1400 MHz 
(J2238+5903) has not been detected 
in radio. However, a nondetection does not rule out the presence of low-level 
radio emission.

Considering the uncertainties of $\zeta$ and $\alpha$, neither model has 
pulsars that are expected to be RQ at 1400 MHz, but RL at 150 MHz.
This result is not surprising as the widening of the beam 
(from 16.8$^{\circ}$ to 22.5$^{\circ}$) is smaller than the uncertainties of 
$\zeta$ and $\alpha$
($\ge10^{\circ}$), a pulsar can at most change from RQ (at 1400 MHz) to RF (at 150 MHz), 
or from RF to RL.
This does indeed happen for two pulsars, using the respective preferred models, for which we thus expect 
a more favorable geometry at 150 MHz:
J0633+0632 should be RL at 150 MHz, but RF at 1400 MHz, and thus it is a good candidate for reobservation
at 150 MHz.
The pulsar J2032+4127 (RL at 150 MHz, but RF at 1400 MHz) has been detected at 2000 MHz, 
and should geometrically be observable at 150 MHz. 
Our nondetection at 150 MHz gives useful constraints for the spectral index (Section \ref{sec-results}).
For this reason, this pulsar is another good candidate for reobservation at 150 MHz.

In Table \ref{tab-pulsars-radioflag},
most pulsars are labeled as RQ and are thus not expected to 
be detectable in radio.
Our nondetections are thus compatible with an unfavorable viewing geometry.
In addition, 
we saw in Section \ref{sec-beam-widening} that RFM broadening is not observed for 
all pulsars, decreasing the number of expected detections.
Finally, in Section \ref{sec-spectral-index} we have discussed 
why $S_{1400} \propto L_r/d^2$ might simply be below
the FR606 HBA sensitivity, even if the beam sweeps Earth.
Combining these factors, the nondetection of all 27 pulsars at 150 MHz 
is compatible with an average spectral index and with RFM-like 
beam-widening for most pulsars.

\section{Conclusion}
\label{sec-conclusions}

We have followed up on 27 radio-quiet gamma-ray pulsars at low radio frequencies (110--190 MHz). 
No pulsed radio emission was detected.
We have established stringent upper limits on their low-frequency radio flux density,
which are considerably more constraining than previous limits in comparable frequency ranges.

Despite the beam-widening 
expected at low radio frequencies, the comparison to simulations shows 
that 
most of our nondetections 
can be explained
by an unfavorable viewing geometry;
for the remaining observations 
(especially those of pulsars detected at higher frequencies),
the sensitivity of our setup was not sufficient.

Based on our geometrical simulations and flux density upper limits,
our best candidates for follow-up observations at 150 MHz are 
J0633+0632 and J2032+4127.
It should be noted that follow-up observations of pulsars detected 
in radio, even if only at a higher frequency,
are easier to analyze as no search for the correct DM value is required.
With this in mind, all five pulsars already detected in radio 
should be reobserved at 150 MHz, potentially except for J1907+0602 for which a low-frequency detection is only possible if the spectrum turns out to be exceptionally steep.

Observations at even lower frequencies ($<100$ MHz) could be performed with NenuFAR
\citep{Zarka18ursi,Zarka20ursi,Bondonneau20nenu}. However, flux extrapolations
will have to take the likely spectral turnover into account.

\begin{acknowledgements}

LOFAR, the Low-Frequency Array designed and constructed
by ASTRON, has facilities in several countries, that
are owned by various parties (each with their own funding
sources), and that are collectively operated by the International
LOFAR Telescope (ILT) foundation under a joint scientific
policy. 
Nan\c{c}ay Radio Observatory is operated
by Paris Observatory, associated with the French Centre
National de la Recherche Scientifique and Universit\'{e}
d'Orl\'{e}ans.

We acknowledge the use of the Nançay Data Center computing facility 
(CDN - Centre de Données de Nançay). 
The CDN is hosted by the Station de Radioastronomie de Nançay in 
partnership with Observatoire de Paris, Université d'Orléans, 
OSUC and the CNRS. The CDN is supported by the Region Centre Val de 
Loire, département du Cher.

The \textit{Fermi} LAT Collaboration acknowledges generous 
ongoing support from a number of agencies and institutes that have 
supported both the development and the operation of the LAT as well 
as scientific data analysis.
These include the National Aeronautics and Space Administration and the
Department of Energy in the United States, the Commissariat \`a 
l'Energie Atomique and the Centre National de la Recherche Scientifique / 
Institut National de Physique Nucl\'eaire et de Physique des Particules in France, 
the Agenzia Spaziale Italiana and the Istituto Nazionale di Fisica Nucleare in Italy, 
the Ministry of Education, Culture, Sports, Science and Technology (MEXT), 
High Energy Accelerator Research Organization (KEK) and Japan Aerospace
Exploration Agency (JAXA) in Japan, and the K.~A.~Wallenberg Foundation, 
the Swedish Research Council and the Swedish National Space Board in Sweden.
Additional support for science analysis during the operations phase is 
gratefully acknowledged from the Istituto Nazionale di Astrofisica in 
Italy and the Centre National d'\'Etudes Spatiales in France. 
This work performed in part under DOE Contract DE-AC02-76SF00515. 
Work at the Naval Research Laboratory is supported by NASA.

\end{acknowledgements}


\bibliographystyle{aa}

\onecolumn

\begin{appendix}
\begin{landscape}

\section{List of radio observations}\label{sec-appendix}

\begin{flushleft}

Table \ref{tab-pulsars-observations} List of all previous radio observations for our 27 targets. 

The (upper limit) spectra of Figure \ref{fig-upperlimits} are based on the data in this table,
as are columns 6--9 in Table \ref{tab-pulsars}.
For detected pulsars, 
Figure \ref{fig-upperlimits} and Table \ref{tab-pulsars} both
use the nominal values (assuming $W/P=0.1$).
Column 1: pulsar name. Column 2: observing frequency. Column 3: pulsar detected in observation? (``y'' if detected, blank otherwise).
Column 4: mean flux density or upper limit. For uniformity, this is the equivalent flux density (assuming $W/P=0.1$).
          When available, the measured flux density (with the measured values of $P$ and $W$) is given in a footnote.
Column 5: telescope name.
Column 6: frequency bandwidth of the observation.
Column 7: observing time $t$ (for observations of this work, we give the effective observing time $t_\text{obs}^\text{eff}$  instead, 
          see eq. (\ref{eq-teff}).
Column 8: minimum S/N assumed by the authors.
Column 9: duty cycle ($W/P$) assumed by the authors.
Column 10: references.       
\end{flushleft}   

\vspace{5ex}

\addtocounter{table}{-1}\addtocounter{table}{-1}
\topcaption{top}
\bottomcaption{This is bottom caption}
\tablecaption{List of radio observations, sorted by pulsar name, observing frequency, and publication date.\label{tab-pulsars-observations}
Notes: $^1$: flux density upper limit calculated as 5 times the 1 $\sigma$ image noise given by \citet{Frail16}, their Table 3.
$^2$: their periodic search.
$^3$: this paper combines new data with previous data from \citet{Maan14}
$^4$: the upper limit resulting from their imaging data represents an averaged flux density (averaged over 16.1 s).
$^5$: nominal flux density limit $S=0.005$ mJy (assuming $W/P=0.1$); detected flux density $S=0.0034$ mJy (with $W/P\sim0.03$).
$^6$: nominal flux density limit $S=0.050$ mJy (assuming $W/P=0.1$); detected flux density $S=0.120$ mJy.
$^7$: nominal flux density limit $S=0.030$ mJy (assuming $W/P=0.1$); detected flux density $S=0.020$ mJy (with $W/P\sim0.02$).
$^8$: nominal flux density limit.
}
\tablefirsthead{
\toprule pulsar& {$f$} & detect? & $S$ 
& telescope & $\Delta f$ & $t$ or $t_\text{obs}^\text{eff}$  & $S/N_\text{min}$ & $W/P$ & reference \\ 
& [MHz] &  & [mJy] &  & [MHz] & [h]  & &  &  & \\ 
\midrule}
\tablehead{%
\multicolumn{2}{c}%
{{\bfseries  Continued from previous page}} \\
\toprule pulsar& {$f$} & detect? & $S$ 
& telescope & $\Delta f$ & $t$ or $t_\text{obs}^\text{eff}$  & $S/N_\text{min}$ & $W/P$ & reference \\ 
& [MHz] &  & [mJy] &  & [MHz] & [h]  & &  &  & \\ 
\midrule}
\tabletail{%
\midrule \multicolumn{2}{r}{{Continued on next page}} \\ \midrule}
\tablelasttail{%
\\\midrule
\multicolumn{2}{r}{{}} \\ \bottomrule}
\begin{xtabular*}{\textwidth}{lccclccccll}
\hline \\
 J0002+6216    & 2000   & y & 0.013$^8$         & GBT         & 700 & 0.47 & 5 & 0.1 & \citet{Clark17}, \citet{Wu18}, J. Wu (personal communication) \\
                   & 1400       & y & 0.022$^8$         & Effelsberg  & 240 & 2    & 5 & 0.1 & \citet{Clark17}, \citet{Wu18}, \citet{WuPHD18} \\ 
               & 150    &   & $<$2.1            & LOFAR/FR606 & 78.125 & 6.5 & 5 & 0.1 & this work       \\ 
 J0007+7303    & 2000   &   & $<$0.006          & GBT & 700 & 2.78 & 5 & 0.1 & \citet{Ray11} \\
                   & 820        &   & $<$0.012          & GBT &  48 & 19.6 & 5 & 0.1 & \citet{Ray11}       and references therein \\ 
                   & 150        &   & $<$11                 & GMRT & 16.7 & 0.25 & 5$^1$ & n/a$^4$ & \citet{Frail16} \\              
               & 150    &   & $<$1.7            & LOFAR/FR606 & 78.125 & 5.4 & 5 & 0.1 & this work       \\ 
 J0106+4855    & 1400   &   & 0.031                 & Effelsberg  & 140 & 0.75 & 5 & 0.1 & \citet{Pletsch12_744} \\
                       & 820    & y & 0.020$^7$         & GBT         & 200 & 0.75 & 5 & 0.1 & \citet{Pletsch12_744} \\ 
                       & 820    & y & 0.020$^7$         & GBT         & 200 & 0.75 & 5 & 0.1 & \citet{Pletsch12_744} \\ 
                       & 350    &   & $<$0.136                  & GBT         & 100 & 0.53 & 5 & 0.1 & \citet{Pletsch12_744} \\
                   & 150        &   & $<$17                         & GMRT & 16.7 & 0.25 & 5$^1$ & n/a$^4$ & \citet{Frail16} \\
                   & 150        &   & $<$1.3                & LOFAR/FR606 & 78.125 & 10.4 & 5 & 0.1 & this work &         \\ 
 J0357+3205    & 350    &   & $<$0.134              & GBT     & 100 & 0.5 & 5 & 0.1 & \citet{Ray11}       &       \\ 
                       & 327    &   & $<$0.043              & Arecibo &  50 &   2 & 5 & 0.1 & \citet{Ray11}              \\ 
                   & 150        &   & $<$13.5               & GMRT & 16.7 & 0.25 & 5$^1$ & n/a$^4$ & \citet{Frail16} \\
               & 150    &   & $<$0.8                & LOFAR/FR606 & 78.125 & 8.9 & 5 & 0.1 & this work &   \\ 
                       & 34         &   & $<$34                     & Gauribidanur& 1.53 & 12 & 5 & 0.1 & \citet{Maan14}$^2$\\ 
 J0359+5414    & 2000   &   & $<$0.018              & GBT         & 700  & 0.67 & 5 & 0.1 & \citet{Clark17}, \citet{Wu18}                \\ 
                       & 2000   &   & $<$0.042              & GBT         & 700  & 0.12 & 5 & 0.1 & \citet{Clark17}, \citet{Wu18}                \\ 
                       & 1400   &   & $<$0.034              & Effelsberg  & 240  & 0.53 & 5 & 0.1 & \citet{Clark17}, \citet{Wu18}                \\ 
                       & 1400   &   & $<$0.023              & Effelsberg  & 240  & 1    & 5 & 0.1 & \citet{Clark17}, \citet{Wu18}                \\ 
                       & 1400   &   & $<$0.015              & Effelsberg  & 240  & 1.92 & 5 & 0.1 & \citet{Clark17}, \citet{Wu18}                \\ 
                   & 150        &   & $<$1.7                & LOFAR/FR606 & 78.125 & 10.5 & 5 & 0.1 & this work &         \\ 
 J0554+3107    & 1400   &   & $<$0.066              & Effelsberg  & 260 & 1 & 5 & 0.1 & \citet{Pletsch13}         \\ 
               & 150    &   & $<$1.2                & LOFAR/FR606 & 78.125 & 7.6 & 5 & 0.1 & this work &   \\ 
 J0622+3749    & 1400   &   & $<$0.030              & Effelsberg  & 250    & 0.53 & 5 & 0.1 & \citet{Pletsch12_744}              \\ 
                   & 1400       &   & $<$0.053              & Effelsberg  & 250    & 0.17 & 5 & 0.1 & \citet{Pletsch12_744}              \\
                       & 1400   &   & $<$0.022              & Effelsberg  & 250    & 0.87 & 5 & 0.1 & \citet{Pletsch12_744}              \\
                       & 1400   &   & $<$0.022              & Effelsberg  & 250    & 0.92 & 5 & 0.1 & \citet{Pletsch12_744}              \\
                       & 820    &   & $<$0.032              & GBT         & 200    & 0.75 & 5 & 0.1 & \citet{Pletsch12_744}              \\
                       & 820    &   & $<$0.032              & GBT         & 200    & 0.75 & 5 & 0.1 & \citet{Pletsch12_744}              \\
                   & 350        &   & $<$0.131              & GBT         & 100    & 0.53 & 5 & 0.1 & \citet{Pletsch12_744}              \\
                       & 150    &   & $<$16.5               & GMRT & 16.7 & 0.25 & 5$^1$ & n/a$^4$ & \citet{Frail16} \\
               & 150    &   & $<$1.4                & LOFAR/FR606 & 78.125 & 5.6 & 5 & 0.1 & this work &   \\ 
 J0631+0646    & 1510   & y & \nodata               & Arecibo     & 300    & 1.15 & 5 & 0.1 & \citet{Clark17}, \citet{Wu18}              \\ 
                       & 1400   & y & 0.018$^8$         & Effelsberg  & 240    & 2    & 5 & 0.1 & \citet{Clark17}, \citet{Wu18}, J. Wu (personal communication)             \\ 
                       & 327    & y & 0.3$^8$           & Arecibo     & 68     & 1.25 & 5 & 0.1 & \citet{Clark17}, \citet{Wu18}, J. Wu (personal communication)            \\ 
               & 150    &   & $<$3.2                & LOFAR/FR606 & 78.125 & 2.0 & 5 & 0.1 & this work &   \\ 
 J0633+0632    & 1510   &   & $<$0.003              & Arecibo     & 300 & 1.17 & 5 & 0.1 & \citet{Ray11}          \\ 
                       & 430    &   & $<$0.052              & Arecibo     &  40 & 1.17 & 5 & 0.1 & \citet{Ray11}                 \\ 
                       & 327    &   & $<$0.075              & Arecibo     &  50 & 0.83 & 5 & 0.1 & \citet{Ray11}                 \\ 
                       & 150    &   & $<$22                         & GMRT & 16.7 & 0.25 & 5$^1$ & n/a$^4$ & \citet{Frail16} \\
               & 150    &   & $<$2.3                & LOFAR/FR606 & 78.125 & 2.0 & 5 & 0.1 & this work &   \\ 
                       & 34             &   & $<$28                         & Gauribidanur& 1.53 & 22.5 & 5 & 0.1 & \citet{Maan14}$^2$ \\
                       & 34             &   & $<$19                         & Gauribidanur& 1.53 & 49.5 & 5 & 0.1 & \citet{Maan15}$^3$ \\
 J1836+5925    & 820    &   & $<$0.010              & GBT         &  48 & 24 & 5 & 0.1 & \citet{Ray11}            \\ 
                       & 350    &   & $<$0.070              & GBT         & 100 &  2 & 5 & 0.1 & \citet{Ray11}           \\ 
                       & 150    &   & $<$18                         & GMRT & 16.7 & 0.25 & 5$^1$ & n/a$^4$ & \citet{Frail16} \\
               & 150    &   & $<$2.5                & LOFAR/FR606 & 78.125 & 2.1 & 5 & 0.1 & this work &   \\ 
                       & 34             &   & $<$55                         & Gauribidanur & 1.53 & 16.5 & 5 & 0.1 & \citet{Maan14}$^2$ \\
 J1838$-$0537  & 2000   &   & $<$0.009              & A & B & C & D & E & \citet{Pletsch12_755}           \\ 
                       & 800    &   & $<$0.082              & A & B & C & D & E & \citet{Pletsch12_755}           \\ 
                       & 150    &   & $<$45.5               & GMRT & 16.7 & 0.25 & 5$^1$ & n/a$^4$ & \citet{Frail16} \\
               & 150    &   & $<$26                         & LOFAR/FR606 & 78.125 & 0.6 & 5 & 0.1 & this work &          \\ 
 J1846+0919    & 1510   &   & $<$0.004              & Arecibo & 300 & 1 & 5 & 0.1 & \citet{SazParkinson10}                \\ 
                       & 350    &   & $<$0.272              & GBT     & 100 & 0.5  & 5 & 0.1 & \citet{SazParkinson10}               \\ 
                       & 350    &   & $<$0.209              & GBT     & 100 & 0.85 & 5 & 0.1 & \citet{SazParkinson10}               \\ 
                       & 150    &   & $<$30.5               & GMRT    & 16.7 & 0.25 & 5$^1$ & n/a$^4$ & \citet{Frail16} \\
               & 150    &   & $<$7.6                & LOFAR/FR606 & 78.125 & 0.9 & 5 & 0.1 & this work &   \\ 
 J1906+0722    & 1400   &   & $<$0.021              & Effelsberg  & 150 & 2 & 5 & 0.1 & \citet{Clark15}           \\ 
               & 150    &   & $<$8.0                & LOFAR/FR606 & 78.125 & 3.0 & 5 & 0.1 & this work &   \\ 
 J1907+0602    & 1510   & y & 0.005$^5$         & Arecibo     & 300 & 0.92 & 5 & 0.1 & \citet{Abdo10_711}, \citet{Ray11} \\ 
                       & 1400   &   & $<$0.022              & Arecibo     & 100 & 0.5 & 5 & 0.1 & \citet{Ray11}          \\ 
                       & 150    &   & $<$40.5               & GMRT & 16.7 & 0.25 & 5$^1$ & n/a$^4$ & \citet{Frail16} \\
                   & 150        &   & $<$7.1                & LOFAR/FR606 & 78.125 & 3.2 & 5 & 0.1 & this work &          \\ 
 J1932+1916    & 1400   &   & $<$0.075              & Effelsberg  & 260 & 1 & 5 & 0.1 & \citet{Pletsch13}                 \\ 
                       & 150    &   & $<$26.5               & GMRT & 16.7 & 0.25 & 5$^1$ & n/a$^4$ & \citet{Frail16} \\
                   & 150        &   & $<$2.9                & LOFAR/FR606 & 78.125 & 4.1 & 5 & 0.1 & this work &          \\ 
 J1954+2836    & 1510   &   & $<$0.007              & Arecibo     & 300 & 0.33 & 5 & 0.1 & \citet{SazParkinson10}                 \\ 
                       & 1510   &   & $<$0.004              & Arecibo     & 300 & 0.75 & 5 & 0.1 & \citet{SazParkinson10}                \\ 
                       & 150    &   & $<$14                         & GMRT        & 16.7 & 0.25 & 5$^1$ & n/a$^4$ & \citet{Frail16} \\
               & 150    &   & $<$2.1                & LOFAR/FR606 & 78.125 & 8.2 & 5 & 0.1 & this work &   \\ 
 J1957+5033    & 820    &   & $<$0.025              & GBT & 200 & 1.36 & 5 & 0.1 & \citet{SazParkinson10}                \\ 
                       & 350    &   & $<$0.225              & GBT & 100 & 0.25 & 5 & 0.1 & \citet{SazParkinson10}                 \\ 
                       & 350    &   & $<$0.122              & GBT & 100 & 0.85 & 5 & 0.1 & \citet{SazParkinson10}                 \\ 
                       & 150    &   & $<$18.5               & GMRT & 16.7 & 0.25 & 5$^1$ & n/a$^4$ & \citet{Frail16} \\
               & 150    &   & $<$1.3                & LOFAR/FR606 & 78.125 & 8.6 & 5 & 0.1 & this work &   \\ 
 J1958+2846    & 1510   &   & $<$0.005              & Arecibo     & 300 & 0.67 & 5 & 0.1 & \citet{Ray11}          \\ 
                       & 150    &   & $<$16                         & GMRT & 16.7 & 0.25 & 5$^1$ & n/a$^4$ & \citet{Frail16} \\
               & 150    &   & $<$2.0                & LOFAR/FR606 & 78.125 & 8.3 & 5 & 0.1 & this work &   \\ 
 J2017+3625    & 2000   &   & $<$0.01               & GBT         & 700    & 1   & 5 & 0.1 & \citet{Clark17}, \citet{Wu18}               \\ 
                       & 1510   &   & $<$0.005              & Arecibo     & 300    & 0.55& 5 & 0.1 & \citet{Clark17}, \citet{Wu18}               \\ 
                       & 1400   &   & $<$0.034              & Arecibo     & 100    & 0.33& 5 & 0.1 & \citet{Clark17}, \citet{Wu18}               \\ 
                       & 1400   &   & $<$0.017              & Effelsberg  & 240    & 0.25& 5 & 0.1 & \citet{Clark17}, \citet{Wu18}               \\ 
                       & 1398   &   & $<$0.050              & Nancay      & 128    & 1.08& 5 & 0.1 & \citet{Clark17}, \citet{Wu18}               \\ 
                       & 1398   &   & $<$0.058              & Nancay      & 128    & 0.87& 5 & 0.1 & \citet{Clark17}, \citet{Wu18}               \\ 
                       & 820    &   & $<$0.043              & GBT         & 200    & 0.75& 5 & 0.1 & \citet{Clark17}, \citet{Wu18}               \\ 
                       & 327    &   & $<$0.17               & Arecibo     &  68    & 0.25& 5 & 0.1 & \citet{Clark17}, \citet{Wu18}               \\ 
                       & 327    &   & $<$0.17               & Arecibo     &  68    & 0.25& 5 & 0.1 & \citet{Clark17}, \citet{Wu18}               \\ 
                       & 327    &   & $<$0.113              & Arecibo     &  68    & 0.47& 5 & 0.1 & \citet{Clark17}, \citet{Wu18}               \\ 
               & 150    &   & $<$1.3                & LOFAR/FR606 & 78.125 & 8.1 & 5 & 0.1 & this work &   \\ 
 J2021+4026    & 2000   &   & $<$0.011              & GBT         & 700 & 1 & 5 & 0.1 & \citet{Ray11}             \\ 
                       & 820    &   & $<$0.051              & GBT         &  48 & 4 & 5 & 0.1 & \citet{Ray11}            \\              
                       & 820    &   & $<$0.053              & GBT         &  48 & 4 & 5 & 0.1 & \citet{Ray11}            \\ 
                       & 150    &   & $<$195                & GMRT & 16.7 & 0.25 & 5$^1$ & n/a$^4$ & \citet{Frail16} \\
               & 150    &   & $<$3.2                & LOFAR/FR606 & 78.125 & 10.3 & 5 & 0.1 & this work &          \\ 
                       & 34             &   & $<$92                         & Gauribidanur& 1.53 & 12 & 5 & 0.1 & \citet{Maan14}$^2$ \\
 J2028+3332    & 2000   &   & $<$0.015              & GBT         & 700  & 0.5 & 5 & 0.1 & \citet{Pletsch12_744}                 \\ 
                       & 1510   &   & $<$0.004              & Arecibo     & 300  & 0.75 & 5 & 0.1 & \citet{Pletsch12_744}                \\
                       & 820    &   & $<$0.046              & GBT         & 200  & 1 & 5 & 0.1 & \citet{Pletsch12_744}               \\ 
                       & 820    &   & $<$0.033              & GBT         & 200  & 0.75 & 5 & 0.1 & \citet{Pletsch12_744}                \\ 
                       & 820    &   & $<$0.033              & GBT         & 200  & 0.75 & 5 & 0.1 & \citet{Pletsch12_744}                \\ 
                       & 327    &   & $<$0.142              & Arecibo     &  25  & 0.42 & 5 & 0.1 & \citet{Pletsch12_744}                \\ 
                       & 150    &   & $<$19                         & GMRT        & 16.7 & 0.25 & 5$^1$ & n/a$^4$ & \citet{Frail16} \\
                   & 150        &   & $<$1.9                & LOFAR/FR606 & 78.125 & 5.2 & 5 & 0.1 & this work &          \\ 
 J2030+4415    & 1400   &   & $<$0.062              & Effelsberg  & 250  & 0.17 & 5 & 0.1 & \citet{Pletsch12_744}                \\ 
                       & 1400   &   & $<$0.035              & Effelsberg  & 250  & 0.53 & 5 & 0.1 & \citet{Pletsch12_744}                \\ 
                       & 1400   &   & $<$0.023              & Effelsberg  & 250  & 1    & 5 & 0.1 & \citet{Pletsch12_744}                \\ 
                       & 1400   &   & $<$0.023              & Effelsberg  & 250  & 1    & 5 & 0.1 & \citet{Pletsch12_744}                \\ 
                       & 1400   &   & $<$0.035              & Effelsberg  & 140  & 0.75 & 5 & 0.1 & \citet{Pletsch12_744}                \\ 
                       & 820    &   & $<$0.038              & GBT         & 200  & 0.75 & 5 & 0.1 & \citet{Pletsch12_744}                \\ 
                       & 820    &   & $<$0.019              & GBT         & 200  & 3.05 & 5 & 0.1 & \citet{Pletsch12_744}                \\ 
                       & 150    &   & $<$33.5               & GMRT & 16.7 & 0.25 & 5$^1$ & n/a$^4$ & \citet{Frail16} \\
                   & 150        &   & $<$0.019              & LOFAR/FR606 & 78.125 & 10.1 & 5 & 0.1 & this work &         \\ 
 J2032+4127    & 2000   & y & 0.050$^6$         & GBT & 700 & 1 & 5 & 0.1 & \citet{Camilo09}, \citet{Ray11}       \\ 
                       & 150    &   & $<$35                         & GMRT & 16.7 & 0.25 & 5$^1$ & n/a$^4$ & \citet{Frail16} \\
               & 150    &   & $<$2.6                & LOFAR/FR606 & 78.125 & 9.6 & 5 & 0.1 & this work     \\ 
 J2055+2539    & 1510   &   & $<$0.106              & Arecibo     & 300 & 0.5 & 5 & 0.1 & \citet{SazParkinson10}          \\ 
                       & 350    &   & $<$0.124              & GBT         & 100 & 0.67 & 5 & 0.1 & \citet{SazParkinson10}                \\ 
                       & 350    &   & $<$0.11               & GBT         & 100 & 0.85 & 5 & 0.1 & \citet{SazParkinson10}                \\ 
                       & 327    &   & $<$0.085              & Arecibo     &  50 & 0.5 & 5 & 0.1 & \citet{SazParkinson10}                 \\ 
                       & 150    &   & $<$29             & GMRT        & 16.7& 0.25 & 5$^1$ & n/a$^4$ & \citet{Frail16} \\
                   & 150        &   & $<$1.3                & LOFAR/FR606 & 78.125 & 5.3 & 5 & 0.1 & this work &          \\ 
                       & 34             &   & $<$60                         & Gauribidanur& 1.53 & 11 & 5 & 0.1 & \citet{Maan14}$^2$ \\
 J2111+4606    & 1520   &   & $<$0.014              & Lovell      & 200  & 14$\times$1 & 5 & 0.1 & \citet{Pletsch12_744}                 \\ 
                       & 820    &   & $<$0.033              & GBT         & 200  & 1 & 5 & 0.1 & \citet{Pletsch12_744}           \\ 
                       & 150    &   & $<$21.5               & GMRT & 16.7 & 0.25 & 5$^1$ & n/a$^4$ & \citet{Frail16} \\
                   & 150        &   & $<$1.9                & LOFAR/FR606 & 78.125 & 7.2 & 5 & 0.1 & this work &          \\ 
 J2139+4716    & 1400   &   & $<$0.022              & Effelsberg  & 250 & 1    & 5 & 0.1 & \citet{Pletsch12_744}          \\ 
                       & 1400   &   & $<$0.029              & Effelsberg  & 250 & 0.53 & 5 & 0.1 & \citet{Pletsch12_744}         \\ 
                       & 820    &   & $<$0.034              & GBT         & 200 & 0.75 & 5 & 0.1 & \citet{Pletsch12_744}         \\ 
                       & 820    &   & $<$0.034              & GBT         & 200 & 0.75 & 5 & 0.1 & \citet{Pletsch12_744}         \\ 
                       & 350    &   & $<$0.171              & GBT         & 100 & 0.53 & 5 & 0.1 & \citet{Pletsch12_744}         \\ 
                       & 150    &   & $<$17                         & GMRT & 16.7 & 0.25 & 5$^1$ & n/a$^4$ & \citet{Frail16} \\
                   & 150        &   & $<$1.1                & LOFAR/FR606 & 78.125 & 9.9 & 5 & 0.1 & this work &          \\ 
                       & 34             &   & $<$74                         & Gauribidanur& 1.53 & 11.5 & 5 & 0.1 & \citet{Maan14}$^2$ \\
 J2238+5903    & 2000   &   & $<$0.007              & GBT         & 700 & 2 & 5 & 0.1 & \citet{Ray11}             \\ 
                       & 820    &   & $<$0.027              & GBT         & 200 & 1.24 & 5 & 0.1 & \citet{Ray11}                 \\ 
                       & 150    &   & $<$17                 & GMRT & 16.7 & 0.25 & 5$^1$ & n/a$^4$ & \citet{Frail16} \\
                   & 150        &   & $<$1.9                & LOFAR/FR606 & 78.125 & 9.7 & 5 & 0.1 & this work &          \\ 
                       & 34             &   & $<$82                         & Gauribidanur & 1.53 & 11 & 5 & E & \citet{Maan14}$^2$ \\
\end{xtabular*}
\end{landscape}

\twocolumn

\end{appendix}


\end{document}